\documentclass[12pt]{article}%
\usepackage{amsmath}%
\usepackage{txfonts}%
\usepackage{amsfonts}%
\usepackage{amssymb}%
\usepackage{graphicx}
\usepackage{color}
\usepackage{setspace}
\usepackage{esint}
\usepackage[rightcaption]{sidecap}
\usepackage{wrapfig}
\usepackage{lipsum}
\usepackage{caption}
\usepackage{longtable}
\usepackage[T2A,T1]{fontenc}
\usepackage[utf8]{inputenc}
\usepackage[russian,english]{babel}
\usepackage[toc,page]{appendix}

\numberwithin{equation}{section}

\newcommand{\be}{\begin{equation}}
\newcommand{\ee}{\end{equation}}

\usepackage{anysize}
\marginsize{2.5cm}{2.5cm}{1cm}{2cm}
\usepackage{mathptmx}
-1in\relax 

\begin{document}
\LARGE
\begin{center}
Feasibility Study For Hydrogen Producing\\ Colony on Mars.
\end{center}
\normalsize
\vskip1cm
\begin{center}
\begin{tabular}{l}
Mikhail V. Shubov         \\
University of MA Lowell    \\
One University Ave,        \\
Lowell, MA 01854           \\
E-mail: mvs5763@yahoo.com  \\
\end{tabular}
\end{center}

\normalsize
\begin{center}
  \textbf{Abstract}
\end{center}
\begin{quote}\begin{quote}
A technologically mature colony on Mars can produce and deliver at least 1 million tons of liquid hydrogen per year to one or more propellant depots at Low Earth Orbit (LEO).  Production of 1 $kg$ of hydrogen at Marian colony and its delivery to LEO requires an energy expenditure of 1.4 $GJ$ on Mars.  LEO propellant depot contains hydrogen produced on Mars and oxygen produced on the Moon or near-Earth asteroids.  This propellant is used to deliver payload from LEO to many destinations in the Solar System, including Mars.  Delivery of 1 $kg$ payload from LEO to Mars requires an energy expenditure of 3.5 $GJ$ on Mars, Moon, and near-Earth asteroids.  The use of liquid hydrogen produced on Mars to deliver astronauts and payload to Mars ensures exponential bootstrap growth of the Martian colony.  Martian Colony and delivery of millions of tons of liquid hydrogen to LEO is the key to Colonization of the Solar System.
The Martian Colony starts transporting liquid hydrogen to LEO only after it grows to significant size.  It should contain about 20 million tons of steel and 3 million tons of plastic in structures and material, as well as several thousand astronauts.  Prior to that time, LEO hydrogen deposit will be supplied by hydrogen from Lunar poles.
\end{quote}\end{quote}

\section{Introduction}
In this work, we discuss the feasibility of a colony on Mars which produces liquid hydrogen.  Most liquid hydrogen is either transported to a propellant depot on 400 $km$ \textbf{low Earth orbit} (LEO) or used itself as propellant in this transportation.
We describe the stages of payload travel from LEO to a space station $4\cdot10^5\ km$ from Mars and from Mars to LEO fuel depot.  We calculate energy requirements for payload delivery in both directions.

Availability of propellant from a LEO depot greatly reduces the cost of transporting any payload from LEO to any destination within the Solar System.  A large fraction of liquid hydrogen in the LEO propellant depot is used to deliver payloads and astronauts to Mars.  These payloads and astronauts enable the Martian hydrogen-producing colony to grow.  Thus, the colony expands via bootstrap.

Hydrogen is a powerful propellant suitable for use in outer space.  First, it can be used as a fuel in conventional rocket engines with oxygen oxidizer.  Unlike hydrogen, oxygen is widely available on the Moon and near-Earth asteroids.
Second, hydrogen is the best propellant to be used in a thermal thruster.  In a thermal thruster, propellant is heated and expelled through a nozzle.  The propellant can be heated by a resistance heater, a nuclear reactor, or a solar concentrator.

Lunar hydrogen production is likely to precede Martian hydrogen production by two decades or more.
Lunar poles contain at least 430 million tons of ice \cite[p.326]{LunarIce}.  Other sources estimate the ice deposits at lunar poles as 660 million tons \cite[p.368]{LunaResources1}.  Based on the data above, the Moon contains 48 million tons to 73 million tons of easily accessible hydrogen at the Lunar poles.
By the time colonization of Mars starts, the Space Industry is mature enough to use at leat a million tons of lunar hydrogen.  Thus, depots and tankers carrying tens or hundreds of thousands of tons of liquid hydrogen are available by that time.
By the time the Martian colony starts exporting hydrogen to LEO, it is relatively mature.  As we mention in Section { 2.1}, it should contain about 20 million tons of steel structure and machinery, as well as about 2.7 million tons of rubber and plastic.  It should also be a home for thousands or tens of thousands astronauts.

In our opinion, Mars is the best source of hydrogen propellant for the time when propellant demand exceeds few million tons.  As mentioned above, Lunar hydrogen deposits are very limited.  Earth has unlimited hydrogen, but delivering any payload from Earth into an orbit is relatively expensive.

Hydrogen propellant production and subsequent relatively inexpensive travel to different destinations within the Solar System enables Humankind to start the colonization of the Solar System.
The Solar System contains almost unlimited resources, which can support a civilization much more populous and advanced than the modern one.  The most important resource for modern industry and civilization is energy \cite{kard, E1}.  The solar energy in space is almost limitless -- the Sun's thermal power is $3.86 \cdot 10^{26}\ W$ \cite[p.14-2]{crc}.  A future civilization, which would harvest 10\% of that power with 10\% efficiency, will have energy production of $3.4 \cdot 10^{25}\ kWh/year$.  The scheme for harvesting this energy is called the Dyson Sphere \cite{Dyson}.  The Dyson Sphere would consist of a multitude of solar energy harvesting space stations orbiting the Sun.  The current global energy production in all forms is $7 \cdot 10^{13}\ kWh/year$ \cite{wes} -- 500 billion times less.  The Asteroid Belt contains almost unlimited material resources -- about $3 \cdot 10^{18}\ tons$ of material composed of metal silicates, carbon compounds, water, and pure metals \cite{ABM}.  Most asteroids are of a carbonaceous type \cite{Asteroids}.  Carbon is very useful for production of food for space travelers, production of fuel for propulsion within space, and for production of plastics for space habitat structures.  High quality steel is also an abundant resource in space, e.g., asteroid 16 Psyche contains $10^{16}\ tons$ of nickel-rich steel \cite{Psycho}.  It has been calculated that Solar System resources can easily sustain a population a million times greater than the global population of today \cite{skymine}.  Thus, colonization of the Solar System will be an extraordinary important step for a development of the civilization.

In Section 2,      we describe the proposed Earth-Mars route.
In Subsection 2.1, we describe the significance of this route in colonization of the Solar System.
In Subsection 2.2, we describe the stages and $\triangle v$ requirements of travel from LEO to a space station $4\cdot10^5\ km$ from Mars.
In Subsection 2.3, we describe the stages and $\triangle v$ requirements of travel from Martian surface to LEO.
In Subsection 2.4, we calculate energy requirements of producing liquid hydrogen on Mars and delivering it to LEO.
In Subsection 2.5, we calculate energy requirements of transporting payload from LEO to Mars.

In Section 3,      we describe electrochemical processes for propellant production on Mars, Moon, and asteroids.
In Subsection 3.1, we describe electric energy production on Mars.
In Subsection 3.2, we describe propellant production on Mars.
In Subsection 3.1, we describe oxygen production on the Moon and asteroids.

In Section 4,      we calculate $\triangle v$ for several space maneuvers.
In Subsection 4.1, we describe Hohmann orbital transitions.
In Subsection 4.2, we calculate $\triangle v$ for orbital transitions in Earth and Mars orbits.
In Subsection 4.3, we calculate $\triangle v$ for orbital transitions in solar orbits.
In Subsection 4.4, we calculate $\triangle v$ for orbital transitions in hybrid orbits.

In Section 5,      we present the conclusion.
In Section 6,      we describe the remaining problems.

\section{Proposed Earth -- Mars route}
\subsection{The role of Earth -- Mars route in Solar System Colonization}
The primary role of the colony on Mars is to produce liquid hydrogen and to deliver it to a propellant depot on 400 $km$ LEO.  Production of liquid hydrogen is energetically expensive.  Delivery of liquid hydrogen from Martian surface to 400 $km$ LEO is much more energetically expensive.

Liquid hydrogen is one of the most powerful propellants which can be used in outer space.  It can be used in chemical rockets with liquid oxygen oxidizer.  It can be used by itself in thermal propulsion rockets.  A liquid hydrogen depot at 400 $km$ LEO enables payloads and astronauts to be delivered from LEO to most destinations within the Solar System.

One of the primary destinations of payloads and astronauts from LEO is Mars. On Mars, the astronauts use the payload brought from Earth to build electric power stations.  Astronauts also build plants for producing propellant from water, carbon dioxide, and electric energy.

As it follows from the above paragraphs, the Martian colony and Earth -- Mars route experience bootstrap growth.  Bootstrap growth is exponential.  Below, we calculate growth coefficient.  Suppose delivery of 1 $kg$ of payload from LEO to Mars requires electric energy $\mathcal{E}_{_{\text{Mars}}}$ to be expanded on Mars.  Energy $\mathcal{E}_{_{\text{Mars}}}$ includes production of propellant and containers on Mars and their transportation to LEO.
Suppose, 1 $kg$ of payload on Mars can be used to build a power station generating electrical power of $\mathcal{P}_{_{\text{Mars}}}$.  Obviously, no power station can have a mass of 1 $kg$.  Construction of a power station requires many astronauts and many tons of payload.  Thus, $\mathcal{P}_{_{\text{Mars}}}$ is the quotient of the electric station power and the mass of payload and astronauts required to build it.

Not all power generated on Mars is for manufacture and transportation of liquid hydrogen to 400 $km$ LEO.  The fraction of power used in this way is denoted $f_{_{\text{Power}}}<1$.
Not all payload exiting LEO is used for construction of power stations on Mars.  Some payload is used to construct propellant production and liquefaction plants and astronaut habitats.  Some payload is used for scientific and exploratory purposes on Mars.  Some payload has Solar System destination other than Mars.  The fraction of propellant used to deliver payload used for construction of power stations on Mars is $f_{_{\text{Propellant}}}<1$.

Payload used for construction of power stations on Mars is delivered at a rate of
    \be
    \label{2.01}
    \frac{P_{_{\text{Total}}}f_{_{\text{Power}}} f_{_{\text{Propellant}}}}
    {\mathcal{E}_{_{\text{Mars}}}}.
    \ee
This payload enables the growth of power production at the rate of
    \be
    \label{2.02}
    P_{_{\text{Total}}}\
    \frac{\mathcal{P}_{_{\text{Mars}}} f_{_{\text{Power}}} f_{_{\text{Propellant}}}}
    {\mathcal{E}_{_{\text{Mars}}}}.
    \ee
Hence, the growth rate of power production as well as the whole colony is
    \be
    \label{2.03}
    \frac{1}{\mathcal{T}_{_e}}=
    \frac{\mathcal{P}_{_{\text{Mars}}} f_{_{\text{Power}}} f_{_{\text{Propellant}}}}
    {\mathcal{E}_{_{\text{Mars}}}},
    \ee
where $\mathcal{T}_{_e}$ is the time it takes the colony to grow by a factor of $e$.  The doubling time is $\mathcal{T}_{_2}=(\ln 2)\mathcal{T}_{_e}$.  From (\ref{2.03}), we obtain
    \be
    \label{2.04}
    \mathcal{T}_{_2}=
    \frac{\mathcal{E}_{_{\text{Mars}}} \ln 2}
    {\mathcal{P}_{_{\text{Mars}}} f_{_{\text{Power}}} f_{_{\text{Propellant}}}}.
    \ee

There are three ways to minimize the doubling time.  First, minimizing the energy cost $\mathcal{E}_{_{\text{Mars}}}$ of transporting payload to Mars with propellant manufactured on Mars.  Second is maximizing the specific power $\mathcal{P}_{_{\text{Mars}}}$ of electricity generating equipment delivered from Earth.  Both of these ways consist of hundreds or thousands of engineering problems. Third is minimizing the power not spent for making and transporting propellant, as well as minimizing the fraction of payload from LEO which does not go to Mars.  The third step is a matter of preference for the future explorers and entrepreneurs.

If $\mathcal{T}_{_2}$ is unreasonably long (decades or centuries), then the Earth -- Mars route can not play an important role in colonization of the Solar System.  In this case, the energy generated on Mars is better spent in expansion of the Martian base itself rather than delivering liquid hydrogen to LEO.  As mentioned in Section { 2.3} below, this will be the case until a mass driver is built on Mars.
If $\mathcal{T}_{_2}$ is under 10 years, then the hydrogen producing colony on Mars may be a key to colonization of the Solar System.  The value of $\mathcal{T}_{_2}$ is likely to change over time.

We can also estimate the size and power production capacity of Martian colony at which it starts transporting liquid hydrogen to LEO.  As we have mentioned in Introduction, Lunar poles contain at least 48 million tons of easily accessible hydrogen.  Thus, Martian hydrogen-producing colony should start transporting hydrogen to LEO only after it can transport at least one million tons of liquid hydrogen per Earth year.  As we estimate in Subsection { 2.4}, the energy required to produce 1 $kg$ of hydrogen and transport it to Earth is $1.4 \cdot 10^9\ J$.  Hence, Martian colony has to produce $1.4 \cdot 10^{18}\ J$ electric energy per year.  The power required to produce this energy is 45 $GW$.

Estimating the mass of machinery and the number of astronauts at the time when Martian colony starts exporting liquid hydrogen to LEO is beyond the scope of this work.  Nevertheless, we can make a very preliminary order of magnitude estimate.  As we estimate in Subsection { 3.2}, it takes 180 $MJ$ to produce one kilogram of plastic and 25 $MJ$ to produce one kilogram of steel. One ExaJoule(EJ), which is $10^{18}\ J$ of energy can be used to produce 20 million tons of steel and 2.7 million tons of plastic from Martian in situ resources.  Astronauts should number in thousands or tens of thousands.

\subsection{Stages of travel from Earth to Mars orbit}
As most routs, the Earth-Mars route is bidirectional. The first route would accommodate technology and astronauts travelling from the Earth to Mars.  The second route would accommodate fuel tanks, samples and possibly other products from Mars as well as returning astronauts traveling back to the Earth.  The steps of travel of the Earth payloads to Mars are listed below.

The first step for the journey from Earth surface to $4\cdot10^5\ km$ Mars orbit is the launch from Earth surface to 400 $km$ LEO.  Payload is delivered to LEO by three stage \textbf{launch vehicle}.  The first two stages are multiply reusable.  They can perform at least 180 launches per year.
The engines of the third stage of each launch vehicle are also reusable.  These engines and all expensive parts of third stages are accumulated at one or more LEO space stations.  From there these engines and parts are returned to Earth by a \textbf{reentry vehicles}.  These vehicles are designed to withstand reentry heating and to land on Earth.

The second step of the journey to $4\cdot10^5\ km$ Mars orbit is transfer from LEO into a Mars-bound orbit around the Sun.  This transfer is performed by a \textbf{transport vehicle} assembled from components brought from Earth to LEO.  Transport vehicles deliver payload and passengers to Mars.  A transport vehicle consists of a propellant tanker, a rocket engine, and a payload or passenger compartment.
One propellant combination is liquid hydrogen fuel with liquid oxygen oxidizer.  It is unreasonably expensive to bring oxygen from Mars or to launch it from Earth.  Oxygen is generated from regolith on the Moon or near-earth asteroids.
Another propellant is liquid hydrogen which is heated by a nuclear reactor, concentrated solar rays, or a resistance heater.  This propellant has much higher specific impulse, but it requires much more complicated and possibly dangerous technology.

The transport vehicle fires its engines and transfers itself into a Mars-bound trajectory.  It takes a transport vehicle 8.5 months to reach Mars.  Very important payloads may be delivered much faster by nuclear reactor powered rockets.
The overall $\triangle v$ for transfer from 400 $km$ LEO into a Mars-bound deep space orbit is about 3.6 $km/s$.

The third step of the journey is transfer from Mars-bound trajectory to a space station at $4\cdot10^5\ km$ Mars orbit.
When the transport vehicle arrives within the vicinity of Mars, it is in a highly elliptical orbit around the Sun.  It is moving at 2.82 $km/s$ with respect to Mars (See Section {  4.3}).
Transport vehicle approaches Mars to the distance of 400 $km$.  At that point, it is travelling at 5.52 $km/s$ with respect to Mars.  Transport vehicle fires its engines in the direction opposite to its motion and slows down to 4.76 $km/s$.  Now it is in an extremely elliptic orbit around Mars.  The $\triangle v$ for this maneuver is about 0.8 $km/s$.
As the transport vehicle returns to the distance of about a $4 \cdot 10^5\ km$ from Mars, it uses its engines to transfer itself into a circular orbit around Mars with a radius of $4 \cdot 10^5\ km$ .  The $\triangle v$ for this maneuver is about 0.35 $km/s$.  At this point, the transport vehicle is joined to a space station in Far Mars Orbit.  Most payload is transported to Mars, some remains in orbit, and some may be sent back to LEO.
The overall $\triangle v$ for the whole trip on which the transport vehicle uses fuel from Earth LEO depot is
      \be
      \label{2.05}
      \begin{split}
      \triangle v&=
      3.6\ \frac{km}{s}\Big|_{_{\text{LEO to Mars trajectory}}}+
      0.8\ \frac{km}{s}\Big|_{_{\text{Mars trajectory to elliptic Mars orbit}}}\\
      &+0.35\ \frac{km}{s}\Big|_{_{\text{elliptic Mars orbit to $4\cdot10^5\ km$ Mars orbit}}}+0.25\ \frac{km}{s}\Big|_{_{\text{other}}}=5.0\ \frac{km}{s}.
      \end{split}
      \ee

\subsection{Stages of travel from Mars to LEO}
In order to explain logistic of material production on Mars and exportation of that material, we introduce the concept of \textbf{energy cost}.  The energy cost of production is the amount of electric energy necessary to produce propellant or material from in situ resources.  The energy cost of transportation is the amount of electric energy necessary to transport a payload from one orbit or location to another.

During initial stages of Mars colonization, every step of travel from Mars to LEO would be powered by rocket engines.  This mode of transportation is energetically expensive.  Only high value payload and returning astronauts would be transported during that time.  During later stages of Mars colonization, the payload should be delivered from Martian surface into 400 $km$ Mars orbit by an electromagnetic \textbf{mass driver} which will be built on Olympus Mons mountain.  A mass driver is a linear accelerator of macroscopic payloads in which the vehicle is pushed forward by a magnetic force \cite{MassDriver1}. Mass driver technology has been experimentally and theoretically studied for over a century \cite{LinMot}.  Currently, mass drivers \textbf{also known as} linear motors are used to propel magnetic levitation trains \cite{LinMot2}.  Toei Oedo subway line in Tokyo is also driven by linear motors \cite{LinMot3}.

Construction of mass driver on Mars enables more payload to be transported from Mars to Earth.  It will also change the nature of payload.  From that point on, most payload consists of liquid hydrogen propellant delivered to LEO.  Moreover, from that time, the total volume of payloads traveling from Mars to the Earth are at least 10 times greater than from the Earth to Mars.  The steps of travel of the Mars payloads to Earth are described below.

The first step for the journey from Martian surface to LEO is the liftoff to 400 $km$ Mars orbit.  Initially, the payload will be transferred by a multiply reusable \textbf{Mars launch vehicle}.  It should be able to accomplish 180 missions per 365-day year.  Martian years have 687 days.  The theoretical minimum $\triangle v$ for Mars launch to 400 $km$ Mars orbit is 3.7 $km/s$.  Realistic $\triangle v$ should be 4.4 $km/s$.
Mars launch vehicle will have two multiply reusable stages.  Mars launch vehicle will use liquid methane fuel, liquid oxygen oxidizer, oxidizer to fuel ratio of 2, and nozzle expansion ratio of 150.  Such engine has combustion chamber temperature of 2,300 $^o$C, and exhaust velocity of 3.1 $km/s$.

As long as Mars launch vehicle will remain the main mode of orbital launch, this step will remain the most energy-expensive part of the journey from Mars to LEO.  It should take at least 8 $kg$ of propellant to place 1 $kg$ payload into 400 $km$ Mars orbit.  As we show in Subsection { 3.2}, the energy cost of propellant is 35 $MJ/kg$.  Hence, the energy cost of orbital launch of payload from Mars is 280 $MJ/kg$.
A 25\% efficient mass driver delivers payload to Mars orbit at an energy cost of
    \be
    \label{2.06}
    \mathcal{E}_{_{\text{MarsOrbit}}}=
    \frac{v^2}{2 \eta}=
    \frac{\big(4,400\ m/s\big)^2}{2 \cdot 0.25} \approx 40\ \frac{MJ}{kg}.
    \ee
As we see from aforementioned data, rocket launch from Martian surface to orbit is at least seven times as energetically expensive as mass driver launch.

The second step of the journey to LEO is transfer from 400 $km$ Mars orbit into an Earth-bound orbit around the Sun.  This transfer is performed by a \textbf{transport vehicle}.  This vehicle uses liquid hydrogen fuel delivered from Mars.  Prior to construction of mass driver on Mars, it would use liquid oxygen manufactured from regolith of Phobos and Deimos and transferred to the fuel depot at 400 $km$ Mars orbit.  Following the appearance of the mass driver, liquid oxygen is delivered from Mars.

The transport vehicle fires its engines and transfers itself into an Earth-bound trajectory.  It takes the vehicle 8.5 months to reach Earth.  Very important payloads may be delivered much faster by nuclear reactor powered rockets.  The overall $\triangle v$ for transfer from 400 $km$ orbit around Mars into an Earth-bound deep space orbit is about 2.2 $km/s$.

The third step of the journey to LEO is transfer from Earth-bound trajectory to a space station $2 \cdot 10^6\ km$ from the Earth.
When the transport vehicle arrives within the vicinity of Earth, it is in a highly elliptical orbit around the Sun.  It is moving at 2.92 $km/s$ with respect to Earth (See Section {  4.3}).
Transport vehicle approaches Earth to the distance of 400 $km$.  At that point, it is travelling at 11.27 $km/s$ with respect to Earth.  Transport vehicle fires its engines in the direction opposite to its motion and slows down to 10.82 $km/s$.  Now it is in an extremely elliptic orbit around Earth.  The $\triangle v$ for this maneuver is about 0.45 $km/s$.
As the transport vehicle returns to the distance of about a $2 \cdot 10^6\ km$ from Earth, it uses its engines to transfer itself into a circular orbit around Earth with a radius of $2 \cdot 10^6\ km$.  The $\triangle v$ for this maneuver is about 0.45 $km/s$.
At this point, hydrogen propellant is transferred to a deep space propellant depot and other payload and passengers are transferred to the space station.
The overall $\triangle v$ for the whole trip from 400 $km$ Mars orbit to the space station $2 \cdot 10^6\ km$ from Earth is
      \be
      \label{2.07}
      \begin{split}
      \triangle v&=
      2.2\ \frac{km}{s}\Big|_{_{\text{400 $km$ Mars orbit to Earth trajectory}}}+
      0.45\ \frac{km}{s}\Big|_{_{\text{Earth trajectory to elliptic Mars orbit}}}\\
      &+0.45\ \frac{km}{s}\Big|_{_{\text{elliptic Earth orbit to $2\cdot10^6\ km$ Earth orbit}}}+
      0.4\ \frac{km}{s}\Big|_{_{\text{other}}}=3.5\ \frac{km}{s}.
      \end{split}
      \ee

The fourth step of the journey to LEO is transfer from the space station $2 \cdot 10^6\ km$ from the Earth to LEO.  This is done by an \textbf{aerobreaker ship}.  This ship goes into a very long elliptic orbit around the Earth.  As the aerobreaker ship approaches Earth, it uses Earth's atmosphere to decrease its speed and to enter into 400 $km$ LEO.  Speed loss by the use of Earth's or another planet's atmosphere is called \textbf{aerobreaking}.
At that orbit, the aerobreaker ship transfers its payload to a space station.  Hydrogen fuel is transferred into a LEO fuel depot.
Aerobreaker ship returns to the space station $2 \cdot 10^6\ km$ from the Earth to LEO.

\subsection{Energy cost of transportation of liquid hydrogen from Mars to LEO}
In this subsection, we consider the energy cost of transporting liquid hydrogen from Mars to LEO only after the mass driver on Mars has been built.
As mentioned in Subsection { 3.2}, the energy cost of producing liquid oxygen is 2.0 $MJ/kg$ and liquid hydrogen is 250 $MJ/kg$.  Based on the energy costs of plastic and steel presented in Subsection { 3.2}, we assume that the energy cost of fuel depot material is 150 $kJ/kg$.  Given that fuel depot is composed of plastic and metal, this is a high estimate.
As we have shown in Eq. (\ref{2.06}), the energy cost of launching a payload into orbit by a mass driver is 40 $MJ/kg$.  Hence, the gross energy cost of materials in 400 $km$ Mars orbit is the following:
liquid hydrogen -- 290 $MJ/kg$, fuel depot material -- 190 $MJ/kg$, and liquid oxygen -- 42 $MJ/kg$.
Hydrogen payload is delivered to the fuel depot $2 \cdot 10^6\ km$ from Earth by a transport vehicle.  The transport vehicle engine uses liquid hydrogen fuel and liquid oxygen oxidizer.  It uses an oxidizer to fuel mass ratio of 3:1.  Even though a ratio of 6:1 produces the best exhaust velocity, such ratio also produces unreasonably high combustion chamber temperature.  The exhaust velocity is 4.2 $km/s$.

The $\triangle v$ for the trip from 400 $km$ Mars orbit to the fuel depot $2 \cdot 10^6\ km$ from Earth is 3.5 $km/s$ according to Eq. (\ref{2.07}).  The fraction of transport vehicle gross mass taken up by the propellant is at least
      \be
      \label{2.08}
      f_{_p}=1-\exp \left(-\frac{\triangle v}{v_{_{\text{exhaust}}}} \right)=
      1-\exp \left(-\frac{3.5\ km/s}{4.2\ km/s} \right)=0.57.
      \ee
The transport vehicle uses propellant fraction $f_{_p}=0.6$.  Mass fraction of the vehicle itself is\\ $f_{_v}=0.1$.  Payload mass fraction is $f_{_0}=0.3$.
The energy costs of delivery of liquid propellant to the fuel depot $2 \cdot 10^6\ km$ from Earth are tabulated below:
      \begin{center}
      \begin{tabular}{|l|l|l|l|}
        \hline
        Component           & Mass     & Unit    & Total   \\
                            & Fraction & Cost    & Cost    \\
                            &          & $MJ/kg$ & $MJ/kg$ \\
        \hline
        Payload H$_{_2}$    & 0.3      & 290     & 87      \\
        Transport vehicle   & 0.1      & 190     & 19      \\
        Propellant H$_{_2}$ & 0.15     & 290     & 44      \\
        Propellant O$_{_2}$ & 0.45     & 42      & 18      \\
        \hline
        Total material      &          &         & 168     \\
        Extra (assumed 43\%)&          &         & 82      \\
        Total               &          &         & 240     \\
        \hline
        Total per unit payload &       &         & 800     \\
        \hline
      \end{tabular}
      \captionof{table}{Costs of hydrogen delivery from Mars to deep space depot \label{T.01}}
      \end{center}
The total energy cost of producing liquid hydrogen and transporting it to the fuel depot $2 \cdot 10^6\ km$ from Earth is 800 $MJ/kg$.

The final step in transportation of liquid hydrogen from Mars to LEO is the transportation from the deep space depot to LEO.  As mentioned above, this step is done by an aerobreaker ship.  First, the aerobreaker ship fires its engines and descends into a very highly elliptic orbit.  The orbit approaches the Earth at a distance of 75 $km$ from the surface.  As the aerobreaker skims the Earth's atmosphere, it slows down.  The aerobreaker ship passes through the atmosphere several times.  Then the aerobreaker uses its engines to enter 400 $km$ LEO.  At that point, it transfers its liquid hydrogen payload to LEO propellant depot.  Then the aerobreaker uses remaining fuel to return to the fuel depot $2 \cdot 10^6\ km$ from Earth.

A fully loaded aerobreaker ship carrying liquid hydrogen payload consists of 15\% empty mass, 40\% propellant mass, and 45\% liquid hydrogen payload mass.  The propellant consists of 25\% liquid hydrogen and 75\% liquid oxygen.  Oxygen is obtained from near-Earth asteroids at a cost of 100 $MJ/kg$.
The energy costs of delivery of liquid propellant to the fuel depot in 400 $km$ LEO are tabulated below:
      \begin{center}
      \begin{tabular}{|l|l|l|l|}
        \hline
        Component           & Mass     & Unit    & Total   \\
                            & Fraction & Cost    & Cost    \\
                            &          & $MJ/kg$ & $MJ/kg$ \\
        \hline
        Payload H$_{_2}$    & 0.45     & 800     & 360     \\
        Propellant H$_{_2}$ & 0.1      & 800     & 80      \\
        Propellant O$_{_2}$ & 0.3      & 100     & 30      \\
        \hline
        Total material      &          &         & 470     \\
        Extra (assumed 34\%)&          &         & 160     \\
        Total               &          &         & 630     \\
        \hline
        Total per unit payload &       &         & 1,400   \\
        \hline
      \end{tabular}
      \captionof{table}{Costs of hydrogen delivery to 400 $km$ LEO depot \label{T.02}}
      \end{center}
The overall energy cost of producing hydrogen on Mars and delivering it to LEO should be 1,400 $MJ/kg$.

\subsection{Energy cost of transportation of payload from LEO to Mars}
Payload is transported from LEO to a space station $4\cdot10^5\ km$ from Mars by a transport vehicle.  The transport vehicle uses a conventional rocket engine with hydrogen fuel, oxygen oxidizer, and 3:1 oxidizer to fuel mass ratio.

The energy cost of liquid hydrogen fuel is 1,400 $MJ/kg$.
Below, we estimate the energy cost of liquid oxygen.
Liquid oxygen at deep space depot has an energy cost of 100 $MJ/kg$.
Calculating the energy cost of transporting liquid oxygen from deep space depot to LEO is beyond the scope of this work.  This energy cost should be considerably lower than the energy cost of transporting liquid hydrogen, since liquid oxygen is much more dense and less cryogenic.  In the previous subsection, we have estimated the cost of transporting liquid hydrogen from deep space depot to LEO depot at 600 $MJ/kg$.  We assume the cost of transporting liquid oxygen to LEO depot to be 300 $MJ/kg$.  Hence, the energy cost of liquid oxygen oxidizer is taken as 400 $MJ/kg$.

The $\triangle v$ for the trip from 400 $km$ LEO to the space station $4 \cdot 10^5\ km$ from Mars is 5.0 $km/s$ according to Eq. (\ref{2.05}).  The fraction of transport vehicle gross mass taken up by the propellant is at least
      \be
      \label{2.09}
      f_{_p}=1-\exp \left(-\frac{\triangle v}{v_{_{\text{exhaust}}}} \right)=
      1-\exp \left(-\frac{5.0\ km/s}{4.2\ km/s} \right)=0.72.
      \ee
Mass fraction of the vehicle itself is $f_{_v}=0.08$.  Payload mass fraction is $f_{_0}=0.2$.
The energy costs of delivery of liquid propellant to the space station $4 \cdot 10^5\ km$ from Mars are tabulated below:
      \begin{center}
      \begin{tabular}{|l|l|l|l|}
        \hline
        Component           & Mass     & Unit    & Total   \\
                            & Fraction & Cost    & Cost    \\
                            &          & $MJ/kg$ & $MJ/kg$ \\
        \hline
        Payload             & 0.2      &         &         \\
        Transport vehicle   & 0.08     &         &         \\
        Propellant H$_{_2}$ & 0.18     & 1,400   & 252     \\
        Propellant O$_{_2}$ & 0.54     &   400   & 216     \\
        \hline
        Total material      &          &         & 468     \\
        Extra (assumed 28\%)&          &         & 132     \\
        Total               &          &         & 600     \\
        \hline
        Total per unit payload &       &         & 3,000   \\
        \hline
      \end{tabular}
      \captionof{table}{Costs of hydrogen delivery from Mars to deep space depot \label{T.01}}
      \end{center}
The total energy cost of transporting payload from LEO to the space station $4 \cdot 10^5\ km$ from Mars is 3,000 $MJ/kg$.  Transporting the payload further to Mars should bring the total energy cost up to 3,500 $MJ/kg$.

\section{Electrochemical processes}
\subsection{Electricity production on Mars}

The first choice of energy production on Mars is the heat engine.  Mars has a vast cold reservoir -- polar caps composed of water and carbon dioxide ice.  Dry ice itself can be used as a working fluid \cite{SEng1}.  Used carbon dioxide can be ejected into the atmosphere.
Carbon dioxide heat engines can use different heat sources.  During initial stages of Mars colonization, heat can be provided by nuclear reactors brought from Earth.  During later stages, lower grade heat provided by Martian atmosphere can be used.  One way to harness the heat of the Martian atmosphere is by transporting dry ice to the Equator.  On Martian equator, temperature varies between about -20$^o$C during daytime and about -80$^o$C at night \cite{MCDatabase}.  Another way to harvest atmospheric heat arises once per Martian year on the South Pole.  During the Winter (Southern Summer) season, lasting 1/4 of the year, temperature at South Pole varies between -75$^o$C and -10$^o$C \cite{MCDatabase}.
The aforementioned temperatures are sufficient for a heat source.
Carbon dioxide has vapor pressure of 0.6 $kPa$ at -126$^o$C, 10 $kPa$ at -103$^o$C and 100 $kPa$ at -78.5$^o$C \cite[p.6-100]{crc}.  Ambient pressure on Mars is 0.6 $kPa$.

The second choice of energy production on Mars is high-altitude wind power.  Currently, there are projects for harvesting Jet Stream wind power on Earth \cite{JSW0}.  The author has suggested harvesting jet stream wind power by turbines on moored airships \cite{ShubovJS}.  One problem on Earth is that hydrogen lifting gas is highly reactive with atmosphere.  This is not the case on Mars.
Mars has vast reservoirs of high altitude wind energy.
Powerful winds are observed at altitudes of 10 $km$ to 30 $km$.  They are similar to Jet Stream on Earth.  These winds are mostly at latitudes 40$^o$N to 60$^o$N during Marian winter and mostly at latitudes 40$^o$S to 60$^o$S during Marian summer \cite[p.309]{MarsZonalWind1}.
According to detailed maps of winds in Mars's upper atmosphere generated by Mars Climate Database, at 100 $Pa$ level there are very strong winds.  For about four Martian months, winds exceeding 85 $m/s$ are blowing over several places in Southern Hemisphere, and for another four Martian months, winds exceeding 130 $m/s$ are blowing over several places over the Northern Hemisphere \cite{MCDatabase}.  Thus, during Martian winter, the power density over some places in Northern Hemisphere is at least 3.2 $kW/m^2$, while during Martian summer, the power density over some places in Southern Hemisphere is at least 1.2 $kW/m^2$.  High altitude wind turbines are supported by airships, thus they can be moved freely over the planet.

The third choice of energy production on Mars is solar power.  Bare solar cells can be brought from Earth.  These cells are assembled into modules and panels with plastic manufactured on Mars.  Solar cells themselves can have high power to mass ratio.  Organic solar cells with specific power of 6,300 $W/kg$ should become available in the future \cite{SCell1}.  Presently, Sunpower C60 Solar Cell has 3.55 $W$ power, 7 $g$ mass, and hence specific power of 507 $W/kg$.  This cell is available for \$3.15 in 2021 \cite{CellC60}.  The powers of solar cells are based on insolation for Earth's surface.  Insolation at Mars's orbit is only 43\% that of insolation at Earth's orbit.  The author suggested increasing the output of solar cells by mounting them on airships.  These airships should be close to the North Pole during Summer and close to the South Pole during Winter.  This would enable the cells to harvest solar energy 24 hours each day \cite{ShubovHAS}.

Airships harvesting solar and wind power in Martian atmosphere is subject to ambient temperatures -120$^o$C to -60$^o$C \cite{MCDatabase}.
These airships would have to be made from fluoroplastics, since other plastics become brittle at cryogenic temperatures \cite{CryoPlast3}.
Mars does have fluoride deposits with concentration of up to 3\% \cite{Ftor2}.

\subsection{Propellant and material production on Mars}

The first step of fuel and oxidizer production on Mars is electrolysis of water.
Electric energy consumption for hydrogen from electrolysis of water is about 198 $MJ/kg$ -- considerably higher than the theoretical minimum of 121 $MJ/kg$ \cite{Electrolysis,Electrolysis1}.
Methane is produced by the reaction
      \be
      \label{3.01}
      \text{CO}_{_2}+4 \text{H}_{_2} \to \text{CH}_{_4}
      +2 \text{H}_{_2}\text{O}+3.2 \ \frac{kJ}{g}.
      \ee
Even though the chemical reactor may consume energy, this energy should be more than compensated for by energy released in reaction (\ref{3.01}).  As we see from (\ref{3.01}), it takes 8 $kg$ hydrogen to produce 16 $kg$ methane.  Hence, energy cost of methane gas is 99 $MJ/kg$.  Oxygen can be considered energetically free byproduct.

It takes about 2.0 $MJ/kg$ to liquefy oxygen \cite[p.155]{Cryogen},  1.3 $MJ/kg$ to liquefy methane \cite[p.11]{CryoGas}, and 50 $MJ/kg$ to liquefy hydrogen \cite[p.4531]{HLiq07}.
Based on the data above, the energy costs of fuel and propellant are 2.0 $kJ/kg$ for liquid oxygen, 101 $kJ/kg$ for liquid methane, and 250 $kJ/kg$ for liquid hydrogen.
The propellant used by Mars launch vehicle includes 1/3 mass methane fuel and 2/3 mass oxygen oxidizer held separately.  Based on the data above, the overall cost of this propellant is 35 $MJ/kg$.

At this point, we calculate the energy cost of producing materials.
The energy cost of producing ethylene from carbon dioxide, water, and electricity is 144 $MJ/kg$ \cite{ElectroCh1}.  Plastic can be produced from ethylene.  The energy cost of plastic production should be with 180 $MJ/kg$.

Martian soil contains Fe$_{_2}$O$_3$ at 17\% concentration \cite[p.68]{MarsComp2}.  In order to produce iron and carbon steel, this ore has to be concentrated and smelted with hydrogen.  The smelting reaction is given in Subsection { 3.3}, Eq. (\ref{3.03}) below.  According to reaction (\ref{3.03}), it takes 0.054 $kg$ of hydrogen to smelt 1 $kg$ iron.  Recalling that energy cost of hydrogen is 198 $MJ/kg$, it follows that hydrogen adds 11 $MJ/kg$ to the energy cost of iron.  Other energy expenses in steel production are generally 14 $MJ/kg$ \cite{Steel2}.  Thus, overall energy cost of carbon steel on Mars is 25 $MJ/kg$.

\subsection{Oxygen extraction from Phobos and Deimos, Moon, and near-Earth asteroids}
Oxygen can be extracted from iron bearing rocks via a cycle consisting of the following two steps.
First, water is electrolyzed into hydrogen and oxygen:
      \be
      \label{3.02}
      \text{H}_{_2}\text{O} \to \text{H}_{_2}+\frac{1}{2}\ \text{O}_{_2}.
      \ee
Second, hydrogen is passed over hot iron-bearing rock.  The following reduction reactions occur:
      \be
      
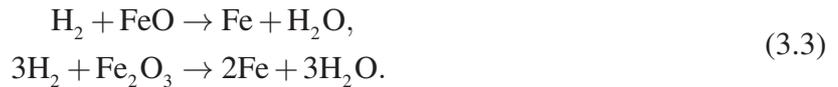
\label{3.03}
      \begin{split}
      \text{H}_{_2}+\text{FeO} &\to \text{Fe}+\text{H}_{_2}\text{O},\\
      3\text{H}_{_2}+\text{Fe}_{_2} \text{O}_{_3} &\to
      2 \text{Fe}+3 \text{H}_{_2}\text{O}.
      \end{split}
      \ee
The net result of the reactions is separation of oxygen from iron oxides.

Below, we estimate the energy cost of oxygen production by the aforementioned process.  First step electrolytically produces 8 $kg$ of oxygen for each 1 $kg$ of hydrogen.  As mentioned in Subsection { 3.2}, the energy cost of electrolytic production of hydrogen is 198 $MJ/kg$.  Hence, the cost of electrolytic production of oxygen is 25 $MJ/kg$.
The second step requires heating about 20 $kg$ rock to about 1,200$^o$C for 1 $kg$ oxygen produced.  About 50\% of the heat is reused.  That step should have energy cost of about 25 $MJ$ per $kg$ oxygen produced.
Overall energy cost of oxygen production from regolith is 50 $MJ/kg$.

\section{Velocity change requirements $\triangle v$ for several space maneuvers}
\subsection{Hohmann orbital transitions}
Let us calculate $\triangle v$ needed for transfer between two orbits around a body with mass $M$.  The radii of the orbits are $r_1$ and $r_2$.  Hohmann \cite{hoh} calculated $\triangle v$ for a two-step transfer.  In the first step, the vehicle is transferred from the circular orbit with radius $r_1$ into an elliptical orbit with the apogee of $r_2$ and perigee $r_1$.  The $\triangle v$ for this step is \cite{hoh}:
    \be
    \label{4.01}
    \triangle v= \sqrt{\frac{M G}{r_1}} \left(\sqrt{\frac{2 r_2}{r_1+r_2}} -1\right),
    \ee
where $G=6.674 \cdot 10^{-11}\ N m^2/kg^2$ is the universal gravitational constant.
In the second step, the vehicle is transferred from the elliptical orbit into the circular orbit with radius $r_2$.  The $\triangle v$ for this step is \cite{hoh}:
    \be
    \label{4.02}
    \triangle v=\sqrt{\frac{M G}{r_2}} \left(1-\sqrt{\frac{2 r_1}{r_1+r_2}}\right).
    \ee
Using Eqs.(\ref{4.01}) and (\ref{4.02}) we calculate the Hohmann velocity change $\triangle v$ requirement for different links of the Earth -- Mars route.

\subsection{Earth and Mars orbits}

The Earth's mass is $5.97 \cdot 10^{24}\ kg$, and its radius is $6.38 \cdot 10^6\ m$.
The Mars's  mass is $6.42 \cdot 10^{23}\ kg$, and its radius is $3.40 \cdot 10^6\ m$
\cite[p.14-3]{crc}.
For LEO 400 km above the Earth's surface, the radius is
$r_{_{\text{E4}}}=6.78 \cdot 10^6\ m$.
For an orbit 400 km above the Mars's surface, the radius is
$r_{_{\text{M4}}}=3.80 \cdot 10^6\ m$.
At these orbits, the velocities of the satellite's rotation are
\be
\label{4.03}
v_{_{\text{E4}}}=\sqrt{\frac{M_{_E} G}{r_{_{\text{E4}}}}}=7.67\ km/s, \qquad
v_{_{\text{M4}}}=\sqrt{\frac{M_{_M} G}{r_{_{\text{M4}}}}}=3.36\ km/s.
\ee

The escape velocity is defined as the velocity necessary to escape into deep space.  It is given by \cite[p.101]{orbmech}:
 \be
 \label{4.04}
 v_{_{\text{escape}}}=
 \sqrt{\frac{2 M G}{r}}.
 \ee
Escape velocities for Earth and Mars are 11.18 $km/s$ and 5.02 $km/s$ respectively \cite[p.14-3]{crc}.
Escape velocities from 400 $km$ above the ground for Earth and Mars are 10.84 $km/s$ and 4.75 $km/s$ respectively.

In order for an object to be shot into free space with velocity $v_f$, it must have velocity $v_{_{\text{escape+}}}$.  This velocity can be calculated by conservation of the total energy $\big(E_{_{\text{total}}}\big)$, which is the sum of the kinetic energy $\big(E_{_{\text{kinetic}}}\big)$ and potential energy $\big(E_{_{\text{potential}}}\big)$:
    \be
    \label{4.05}
    \begin{split}
    v_{_{\text{escape+}}}&=\sqrt{\frac{2}{m} E_{_{\text{total}}}}=
    \sqrt{\frac{2}{m} \left(E_{_{\text{potential}}}+E_{_{\text{kinetic}}}\right)}\\
    &=
    \sqrt{\frac{2}{m} \left(m\frac{M G}{r_1}+\frac{m v_f^2}{2}\right)}=
    \sqrt{\frac{2M G}{r_1}+v_f^2}.
    \end{split}
    \ee

Escaping from 400 $km$ LEO into free space requires
    \be
    \label{4.06}
    \triangle v=v_{_{\text{escape}}}-v_{_{\text{4}}}=
    \left\{
    \begin{array}{ll}
        3.17\ km/s & \text{ for Earth,}\\
        1.97\ km/s & \text{ for Mars.}
    \end{array}
    \right.
    \ee
Escaping from 400 $km$ LEO into free space and retaining free space velocity $v_f$ requires
    \be
    \label{4.07}
    \triangle v=v_{_{\text{escape+}}}-v_{_{\text{4}}}=
    \left\{
    \begin{array}{ll}
        \sqrt{\left(10.84\ km/s\right)^2+v_f^2}-7.67\ km/s & \text{ for Earth,}\\
        \sqrt{\left( 4.75\ km/s\right)^2+v_f^2}-3.36\ km/s & \text{ for Mars.}
    \end{array}
    \right.
    \ee

\subsection{Solar orbits}
The Sun has a mass of $1.99 \cdot 10^{30}\ kg$ \cite{Sun}.  Earth's average distance from the Sun is\\  $r_1=1.50 \cdot 10^{11}\ m$ \cite{AU}, and Mars's average distance from the Sun is $r_2=2.28 \cdot 10^{11}\ m$ \cite[p.14-3]{crc}.
Using formula (\ref{4.01}), we calculate  that $\triangle v$ requirement for transfer from deep space near Earth's orbit to the elliptical orbit with an apogee at Mars's orbit is 2.92 $km/s$.
Using (\ref{4.02}), we calculate  that $\triangle v$ requirement for transfer from Mars's orbit to the elliptical orbit with a perigee at Earth's orbit is $2.82\ km/s$.

\subsection{Hybrid Earth and Solar orbits}
We use (\ref{4.07}) to calculate $\triangle v$ necessary to escape from $400\ km$ LEO into different Solar orbits.  Escaping from $400\ km$ LEO into a Mars bound elliptical orbit requires
 \be
 \label{4.08}
 \triangle v=
 \sqrt{\left(10.84\ km/s\right)^2+\left(2.92\ km/s\right)^2}-7.67\ km/s=
 3.56\ km/s.
 \ee
Escaping from $400\ km$ orbit around Mars into an Earth bound elliptical orbit requires
 \be
 \label{4.09}
 \triangle v=
 \sqrt{\left(4.75\ km/s\right)^2+\left(2.82\ km/s\right)^2}-3.36\ km/s=
 2.16\ km/s.
 \ee

\section{Conclusion}
A colony on Mars which has reached large size and advanced technological maturity can act as a supplier of liquid hydrogen propellant for one or more fuel depot at LEO.  This colony contains at least 20 million tons of steel, 2.7 million tons of plastic, 45 $GW$ electric power generation capacity and thousands or tens of thousands astronauts. It delivers at least a million tons of liquid hydrogen per Earth year to LEO.

Delivery of liquid hydrogen from Mars to LEO consists of at least four steps.  Fist, the liquid hydrogen has to be launched from Marian surface into 400 $km$ orbit around Mars.  This is accomplished by the use of a mass driver located on Olympus Mons mountain.  From there, liquid hydrogen is stored in tankers or fuel depots.  Second, liquid hydrogen is carried by a transport vehicle which leaves the orbit around Mars and goes into an Earth-bound trajectory.  Third, the transport vehicle comes very close to the Earth and uses its engines to slow down and then to enter an orbit of a fuel depot and space station $2 \cdot 10^6\ km$ from Earth.  From the deep space fuel depot, hydrogen is transported to LEO by an aerobreaker ship.  This ship uses Earth's atmosphere to slow down and to enter LEO

Liquid hydrogen in LEO can be used to enable transportation of payloads to many destinations in the Solar System.  A significant stream of payload and astronauts is sent to Mars.  This enables the growth of the colony, which produces hydrogen.  The growth of a Martian colony is a bootstrap process in which liquid hydrogen produced on Mars enables delivery of payload and astronauts to Mars.

In order to deliver 1 $kg$ liquid hydrogen payload to Mars, 1.4 $GJ$ electric energy has to be expanded on Mars.  In order to deliver 1 $kg$ payload from LEO to Mars using liquid hydrogen of Martian origin and oxygen from near Earth asteroids, a total of 3.5 $GJ$ has to be expanded on Mars and asteroids.  Bringing the best equipment to build electric stations on Mars is necessary to ensure rapid growth of the colony.

\section{Remaining problems}
In formulating the remaining problems, we must keep in mind that physical implementation of space projects remains a far yet very important goal.  We do not yet know how far in the future it will be and what technology will exist at that time.  Space colonization may start as early as this decade.  Space colonization may start only at the end of this century or later.  Currently, we can only construct conceptual designs and models based on the state of the art technology.

In order for a working model to exist, much more detailed calculations are necessary for every step of transportation between Earth and Mars.  Moreover, every machine used on the route between Mars and Earth must be developed in much greater detail.  These machines include transport ships for transportation in deep space, Mars launch vehicles, and aerobreaker ships for both Earth and Mars.  They also include the mass driver and electric power stations on Mars.

Technologies for generating the most electric power on Mars with the least amount of material brought from Earth must be developed.  These technologies will enable fast growth of liquid hydrogen producing colony on Mars.

Researchers should also look at advantages and disadvantages of using different technologies on the Earth-Mars route.  We should consider different modes of rocket propulsion: chemical, electric, and thermal.  Novel concepts should be developed in detail.  Hopefully, development of a hydrogen producing colony on Mars is an important step toward colonization of the Solar System.

\end{document}